\documentclass[aoas,nameyear,seceqn,dvips]{arximspdf}
\usepackage{dcolumn}
\usepackage{graphicx}

\doi{10.1214/10-AOAS426}
\volume{5}
\issue{2A}
\pubyear{2011}
\firstpage{1081}
\lastpage{1101}

\makeatletter
\renewcommand{\widehat}{\hat}
\newcolumntype{d}[1]{D{.}{.}{#1}}
\newcommand{\bfX}{\mathbf{X}}
\newcommand{\bfbeta}{\bolds{\beta}}
\newcommand{\GG}{\gg}
\newcommand{\eqref}[1]{(\ref{#1})}
\makeatother

\begin{document}
\begin{frontmatter}

\title{Survival ensembles by the sum of pairwise differences
with application to lung cancer~microarray studies}
\runtitle{Ensembles by the sum of pairwise differences}

\begin{aug}
\author[a]{\fnms{Brent A.} \snm{Johnson}\corref{}\thanksref{t1}\ead[label=e1]{bajohn3@emory.edu}}
\and
\author[a]{\fnms{Qi} \snm{Long}\ead[label=e2]{qlong@emory.edu}}
\runauthor{B. A. Johnson and Q. Long}
\affiliation{Emory University}
\address[a]{Department of Biostatistics\\
Rollins School of Public Health\\
Emory University\\
1518 Clifton Rd. NE\\
Atlanta, Georgia 30322\\
USA \\
\printead{e1}\\
\hphantom{\textsc{E-mail:} }\printead*{e2}} %adresu isvedimo komanda
%gale!

\end{aug}
\thankstext{t1}{Supported in part by a grant from
the National Institutes of Allergies and Infectious Diseases
(R03AI068484) and
Emory's Center for AIDS Research.}

% HISTORY:
\received{\smonth{9} \syear{2009}}
\revised{\smonth{9} \syear{2010}}

% ABSTRACT
%
\begin{abstract}
Lung cancer is among the most common cancers in the United States,
in terms of incidence and mortality. In 2009, it is estimated that
more than 150,000 deaths will result from lung cancer alone.
Genetic information is an extremely valuable data source in characterizing
the personal nature of cancer. Over the past several years, investigators
have conducted numerous association studies where intensive genetic data
is collected on relatively few patients compared to the
numbers of gene predictors, with one scientific goal being to
identify genetic features associated with cancer recurrence or survival.
In this note, we propose high-dimensional survival analysis through a~new application of
boosting, a powerful tool in machine learning.
Our approach is based on an accelerated lifetime model and
minimizing the sum of pairwise differences in residuals.
We apply our method to a recent microarray study of lung adenocarcinoma
and find that
our ensemble is composed of 19 genes, while a proportional hazards (PH) ensemble
is composed of nine genes, a proper subset of the 19-gene panel. In one
of our simulation scenarios,
we demonstrate that PH boosting in a misspecified model tends to
underfit and ignore moderately-sized covariate effects, on average.
Diagnostic analyses suggest that
the PH assumption is not satisfied in the microarray data and may
explain, in part, the discrepancy
in the sets of active coefficients. Our simulation studies and
comparative data analyses
demonstrate how statistical learning by PH models alone is
insufficient.\looseness=-1
\end{abstract}

% KEYWORDS
%

\end{frontmatter}
%

%s1 ###
\section{Introduction}\label{sec:intro}

In 2009, lung (and bronchus) cancer is projected to be
the third most incident cancer site (behind prostate and breast) in the
United States.
The National Cancer Institute
estimates that nearly 220,000 men and women will be
diagnosed with and nearly 160,000 men and women\vadjust{\eject}
will die from lung and bronchus cancer in 2009 (\href{http://seer.cancer.gov}{seer.cancer.gov}).
Large data sets containing
clinical, microarray, miRNA, and other genomic data for virtually all
types of cancer
are available online and one
expects these data sets to multiply in the future. While the demand from
scientific investigators for systematic summary of large data sets is high,
there is a short supply of easy-to-use, out-of-the-box methods for
survival outcomes.
The goal of this note is to propose a new statistical learner for\vadjust{\goodbreak}
survival data
by boosting a~weighted concentration measure
and applying the method to a challenging scientific problem in lung cancer.
The routines developed for this paper complement the \texttt{mboost} package
[Hothorn and B\"{u}hlmann (\citeyear{Hothorn2007})]
in {\texttt R} and are freely available from the first author's website.

Boosting is a ubiquitous concept in machine learning and popular among
statisticians for model fitting, prediction, and variable selection.
While early
applications were driven by problems in classification and
discrimination [Freund
and Schapire (\citeyear{Freund1996,Freund1997}); Breiman
(\citeyear{Br1998})],
it is now
known that boosting applies to a general class of
function estimation problems by stage-wise descent of a well-defined,
convex loss function
[cf. Friedman, Hastie and Tibshirani (\citeyear{Friedman2000});
 B\"{u}hlmann and Hothorn
(\citeyear{B2007})].
In this paper we propose rank-based boosting of survival data in a
semi-parametric
accelerated lifetime or failure time (AFT) model [Cox and Oakes
(\citeyear{Cox1984});
Kalbfleisch and Prentice (\citeyear{Kalbfleisch2002})].
In the linear AFT model, the (natural) logarithm of
the lifetime $T_i$ is related to a $d$-vector of predictors $\bfX
_i=(X_{i1},\ldots ,X_{id})^{\mathrm T}$, that~is,
%
%e1.1 ###
\begin{equation}\label{eq:lin}
\log T_i = \sum_{j=1}^d \beta_jX_{ij} + \varepsilon_i  \qquad  (i=1,\ldots ,n),
\end{equation}
$(\varepsilon_1,\ldots ,\varepsilon_n)$ are random errors from an
unknown common distribution,
and $\bfbeta=(\beta_1,\ldots ,\beta_d)^{\mathrm T}$ is an unknown
coefficient vector to be estimated.
Without loss of generality, we assume that the predictors have been
standardized to
have mean zero and unit variance.
The observed data are $\{(U_i,\Delta_i,\bfX_i), i=1,\ldots ,n\}$, where
$U_i=\min(T_i,C_i)$, $\Delta_i=I(T_i\leq C_i)$,
$C_i$ is a random censoring variable for the $i$th subject,
and $I(\cdot)$ denotes the indicator function. Because the linear AFT
model is based on the linear model,
the coefficients and their estimates have an interpretation familiar to
a broad audience. Indeed, Sir David
Cox highlighted parameter interpretation when describing the appeal of
the AFT model compared to
parameters in hazards regression models [Reid (\citeyear{Reid1994})].
Although boosting in the AFT model has already
been described by Hothorn et al. (\citeyear{Hothorn2006}) via
inverse-probability weighting (IPW), the
method here is based on an entirely different principle and embodies
different assumptions (see Section~\ref{sec:methods}).
In Section~\ref{sec:sim} we compare rank-based
and IPW ensembles to assess how technical assumptions affect
performance in statistical learners.

Ridgeway (\citeyear{Ridgeway1999}) first proposed the idea of boosting
survival data through
Cox's (\citeyear{Cox1972}) partial log-likelihood
in a proportional hazards (PH) model (see Section~\ref{sec:cox}).
In general, adopting the PH model and complementary
partial likelihood analysis is the conventional method in ordinary survival
regression as well as their extensions to model selection and
statistical learning.
However, the proportional hazards assumption may be incorrect.
When the PH assumption is incorrect but
one proceeds with partial likelihood analysis, there can be serious
side effects on classic statistical inference,
resulting in incorrect conclusions [Lin, Wei and Ying (\citeyear
{Lin1993})]. Although the consequences of model misspecification
on variable selection may be subtle for any given data set, a violation
of the PH model assumption
would, for example, obviate the oracle property [Fan and Li (\citeyear
{Fan2002});
Johnson, Lin and Zeng (\citeyear{Johnson2008})] for the relevant
penalized partial likelihood estimators. One may use
standard tools to diagnose the PH assumption in survival data
[cf. Schoenfeld (\citeyear{Schoenfeld1982}); Lin, Wei and Ying
(\citeyear{Lin1993}); Grambsch and Therneau (\citeyear{Grambsch1994})] and if
deviations occur, then one can take corrective action through various
relaxations of the PH model.
A second way to circumvent potential pitfalls in PH model
misspecification is to posit a different
statistical model, such as the AFT model.
In Section~\ref{sec:sim} we compare and contrast
coefficient ensembles across competing survival models using a variety
of performance measures.

This paper was originally motivated by the authors' collaborations with
investigators
who collect and analyze high-dimensional microarray data.
Recently, Morris et al. (\citeyear{Morris2005}) analyzed microarray
data collected from two
studies, one conducted at Harvard University
and another at the University of Michigan. Both studies used Affymetrix
oligonucleotide arrays but used
different versions of Affymetrix chips. Morris et al. pooled the data
using a~``partial probeset'' method to match chip types and then used the
pooled data to identify genes
associated with mortality due to lung adenocarcinoma, a nonsmall cell
form of lung cancer.
The goal of our analysis is to
develop a mortality model of genetic factors for lung adenocarcinoma.
Morris et al. achieved this goal through one-gene-at-a-time Cox PH
models and then controlled for false discovery.
In Section~\ref{sec:results} we perform simultaneous estimation
and variable selection on the same microarray data via boosting.
We find that PH boosting leads to
a nine-gene model and rank-based boosting leads to a 19-gene model,
with the former active set a
proper subset of the latter model. Boosting the same data using IPW
methods leads one to conclude
that 94 genes are active, most of which do not appear in either of the
other two methods.

In addition to our substantive findings in the lung cancer data, we
also make
the following two methodological contributions. In Section~\ref
{sec:compare} we provide an
analysis of nursing home data
where the sample size far exceeds the number of predictors so we can
apply standard PH model diagnostics.
The diagnostic tool fails to support the PH model assumption and
variable selection
using PH versus AFT models leads to rather different sets of active covariates.
Later in the same section, we apply blackbox boosting using rank-based methods
to breast cancer data [Street, Mangasarian and Wolberg (\citeyear{Street1995})]
and compare our results to IPW boosting [Hothorn et al. (\citeyear
{Hothorn2006});
B\"{u}hlmann and Hothorn (\citeyear{B2007})].
This analysis exemplifies key differences between rank-based and IPW
methods even when
one adopts the same AFT model.
To the best of our knowledge, this is the first paper to
execute nonlinear regression in the AFT model using the rank-based
Gehan loss and
regression trees as base learners.\vadjust{\goodbreak}

%s2 ###
\section{Methods}\label{sec:methods}

%s2.1 ###
\subsection{Boosting}\label{ssec:boost.alg}

For many applications, a common goal is to estimate the population minimizer,
%
%e2.1 ###
\begin{equation}\label{eq:f0}
f_0(\cdot) = \operatorname{argmin} E [\rho\{U,\Delta,f(\bfX)\} ],
\end{equation}
where $\rho$ is a convex loss function, differentiable with respect to
$f$, and $f$ is a function to
be estimated in the generalization of model \eqref{eq:lin}, that is,
%
%e2.2 ###
\begin{equation}\label{eq:gen.aft}
\log T_i = f(\bfX_i) + \varepsilon_i,
\end{equation}
and $\varepsilon_i$ were described in Section~\ref{sec:intro}.
We assume that the observed data $\{(U_i,\Delta_i,\bfX_i),  i=1,\ldots
,n\}$
are a random sample of observations from a common distribution function;
hence, an ordinary strong law suggests the expectation on the
right-hand side
of \eqref{eq:f0} is well approximated by a sample average.
Then, the goal of boosting is to minimize the empirical loss function,
that is,\vspace*{-3pt}
%
%e2.3 ###
\begin{equation}\label{eq:rho.sum}
\widehat{f}(\cdot) = \operatorname{argmin} \frac{1}{n}\sum_{i=1}^n \rho\{
U_i,\Delta_i,f(\bfX_i)\}.
\end{equation}
As long as $\rho(\cdot)$ is convex,
we note that the computational exercise in \eqref{eq:rho.sum} is well
defined even
when the population parameter $f_0(\cdot)$ in \eqref{eq:f0} is
awkward or difficult to interpret.

With only minor notation changes, we briefly outline functional
gradient descent (FGD)
as given in Friedman (\citeyear{Friedman2001}) and B\"{u}hlmann and
Hothorn (\citeyear{B2007}):
\begin{longlist}[(S5)]
\item[(S1)] Initialize $\widehat{f}^{[0]}\equiv0$.
\item[(S2)] Increment $m$ by 1 and compute the negative gradient
$-(\partial/\partial f)\rho\{U,\break\Delta, f(\bfX)\}$. Define $Z_i$ as
the evaluation of the
negative gradient at $\widehat{f}^{[m-1]}(\bfX_i)$, that is,
\[
Z_i = -\frac{\partial}{\partial f}\rho\{U_i,\Delta_i,f\}
\bigg|_{f=\widehat{f}^{[m-1]}(\bfX_i)}
\]
for $i=1,\ldots ,n$.
\item[(S3)] Fit the new pseudo data $\{(Z_i,\Delta_i,\bfX
_i), i=1,\ldots ,n\}$ through
a base procedure to form the ensemble update, $\widehat{g}^{[m]}(\cdot)$.
\item[(S4)] Define the ensemble iterate $\widehat{f}^{[m]}(\cdot) =
\widehat{f}^{[m-1]}(\cdot) + \nu\cdot\widehat{g}^{[m]}(\cdot)$,
where $\nu$ is a user-defined step-length factor less than or
equal to 1 and strictly greater than 0.
\item[(S5)] Iterate from steps S2 to S4 for a user-defined number of
iterations, that is, ``$m_{\mathrm{stop}}$.''
\end{longlist}

The final estimate $\widehat{f}$ is driven by the number of iterations
$m_{\mathrm{stop}}$. Hence,
$m_{\mathrm{stop}}$ is a parameter that requires tuning: fewer iterations
lead to simple models but worse
prediction, while increasing $m_{\mathrm{stop}}$ increases model complexity
and eventual overfitting.

%s2.2 ###
\subsection{The Gehan loss function}

We motivate our loss function through coefficient estimation in the
semi-parametric
AFT model \eqref{eq:lin}. In the case where $d=1$, Prentice (\citeyear
{Prentice1978}) proposed linear rank tests for the null
hypothesis that the slope is zero. Tsiatis (\citeyear{Tsiatis1990})
inverted the linear rank tests to
form a class of weighted logrank estimating functions. For a particular choice
of inefficient weight function, the weighted logrank estimating
function reduces to\looseness=-1
%
%e2.4 ###
\begin{equation}\label{eq:EE.Gehan}
n^{-2}\sum_{i=1}^n \sum_{j=1}^n \Delta_i(\bfX_i-\bfX_j)I\{
e_i(\bfbeta)\leq e_j(\bfbeta)\}.
\end{equation}
We note that \eqref{eq:EE.Gehan}
is the $d$-dimensional gradient of the following Gehan-type (\citeyear
{Gehan1965}) loss function:
%
%e2.5 ###
\begin{equation}\label{eq:LG.simp}
- n^{-2}\sum_{i=1}^n\sum_{j=1}^n \Delta_i \{e_i(\bfbeta
)-e_j(\bfbeta)\}I\{e_i(\bfbeta) \leq e_j(\bfbeta)\},
\end{equation}
where $e_i(\bfbeta)=\log U_i - \bfbeta^{\mathrm T}\bfX_i$. If we relax the
restriction that $f(\cdot)$ is a linear predictor, expression \eqref
{eq:LG.simp} is still a proper
convex loss function and reduces to Jaeckel's (\citeyear{Jaeckel1972})
dispersion criterion in uncensored data.
In the context of boosting, we define the Gehan loss as the following
weighted sum of pairwise differences:
%
%e2.6 ###
\begin{eqnarray}\label{eq:LG}
D_G(f) &=& - n^{-2}\sum_{i=1}^n\sum_{j=1}^n \Delta_i (e_i-e_j)I(e_i
\leq e_j)  \nonumber
\\[-8pt]
\\[-8pt]
&=& n^{-1} \sum_{i=1}^n  \Biggl\{-n^{-1}\Delta_i\sum_{j=1}^n (e_i -
e_j)I(e_i \leq e_j)  \Biggr\},
\nonumber
\end{eqnarray}
where $e_i=\log U_i - f(\bfX_i)$. We compute the negative gradient in
step (S2) in Section~\ref{ssec:boost.alg}
as the following difference:
$-(\partial/\partial f) D_G = -(\Gamma_1 - \Gamma_2)/n$, where
$\Gamma_1 = \Delta_i \sum_{j=1}^n I(e_i \le e_j)$
and $\Gamma_2 = \sum_{j=1}^n \Delta_j I(e_i \ge e_j)$.
Now, we see that the definition of $\rho\{U_i,\Delta_i,f(\bfX_i)\}$
from the prototypical boosting algorithm in
Section~\ref{ssec:boost.alg} is the expression in curly brackets on
the right-hand side of \eqref{eq:LG}.

%s2.3 ###
\subsection{Parameter tuning}

We require a criterion whereby we can assess model fit
in terms of error and complexity. Because $D_{G}(f)$ is a convex loss function,
we may use it to simply perform $V$-fold cross-validation (CV). Unless
otherwise specified,
we adopt $5$-fold CV to tune $m_{\mathrm{stop}}$ for all data analyses and
simulation studies below.

As noted by an anonymous referee, boosting is known to be a slow
learner and hence slow in convergence
[cf. B\"{u}hlmann and Yu (\citeyear{B2003}); Blanchard, Lugosi and
Vayatis (\citeyear{Blanchard2004});
Zhang and Yu (\citeyear{Zhang2005})]. This suggests that for a~fixed
step-length factor $\nu$ in step (S4) of the boosting algorithm
described in Section~\ref{ssec:boost.alg},
convergence occurs only after a large number of boosting iterations and
will eventually
overfit if boosting iterates indefinitely. In practice, we found that
convergence was more difficult for
``small'' data sets (i.e., larger number of iterations needed for
large $n$ and small $d$) than for
data sets with large numbers of predictors, although this will be more
closely related to the signal-to-noise
ratio, in general.
For fixed step-length $\nu$, our experience suggests that $V$-fold CV
is quite reliable for parameter tuning.
Where asymptotic analysis suggests allowing $\nu\rightarrow0$, we
adopt B\"{u}hlmann's recommendation
of setting $\nu=0.1$. A simple sensitivity analysis revealed that
coefficient ensembles were rather
insensitive to mild differences in step-length.

%s3 ###
\section{Loglinear vis-a-vis hazards regression for lifetime data
analysis}\label{sec:cox}

$\!\!\!$Rather than assert model \eqref{eq:lin}, a popular alternative for
lifetime data is to model the hazard function,
$\lambda(t, \bfX_i) = \lim_{h\downarrow0}
\operatorname{pr}(t \le T_i <
t+h | T_i\ge t, \bfX_i)$.
Cox's (\citeyear{Cox1972}) proportional hazards (PH) assumption
asserts that
%
%e3.1 ###
\begin{equation}\label{eq:coxph}
\lambda(t,\bfX_i) = \lambda_0(t) \exp(\beta_1X_{i1} + \cdots+
\beta_dX_{id}),
\end{equation}
where $\lambda_0(t)$ is an arbitrary function of time. Because one
models the hazard function in \eqref{eq:coxph},
the coefficients are interpreted on a log relative risk scale.
Regardless of coefficient interpretation, the maximum
partial likelihood estimator minimizes the following convex loss
function:
%
%e3.2 ###
\begin{equation}\label{eq:pl}
-\frac{1}{n}\sum_{i=1}^n \Delta_i  \biggl[ \bfX_i^{\mathrm T}\bfbeta-
\log \biggl\{\sum_{j\dvtx  U_j\ge U_i} \exp(\bfX_j^{\mathrm T}\bfbeta)
\biggr\}  \biggr],
\end{equation}
and coefficient ensembles are constructed accordingly.
Note that when the errors in \eqref{eq:lin} are normally distributed,
the proportional hazards
model in~\eqref{eq:coxph} is misspecified. Similarly, it is easy to
construct distributions where the PH model
is correct
and the loglinear model in \eqref{eq:lin} is incorrect. Both models
are correct only when the distribution of the
lifetime variable is an extreme value. Graphical displays and formal
hypothesis tests for the (in)validity of the PH model
have been a research topic for survival enthusiasts for more than four
decades. The PH model can
fail in one of three ways [cf. Lin, Wei and Ying (\citeyear
{Lin1993})]: (a) the PH assumption, (b)~the functional form of predictors,
and (c) the link function. Any violation can have serious side effects
on partial likelihood inference,
including inefficient coefficient estimates, hypothesis tests with the
wrong size, and confidence intervals with
the wrong coverage [Lin, Wei and Ying (\citeyear{Lin1993})]. In
addition to numerous research papers on detecting deviations from the
PH model, some text books are dedicated to the topic [cf. Therneau and
Grambsch (\citeyear{Therneau2000})] and routine
tools are available in standard software. In our analysis of the
microarray data in Section~\ref{sec:results},
we use a diagnostic tool developed
by Grambsch and Therneau (\citeyear{Grambsch1994}) that investigates
whether the covariate effect is constant over time.

%Model evaluation for survival data is tricky because of multiple
%biases: one potential bias due to
%model misspecification and one bias due to variable selection.
%Where model evaluation in complete data rely on classic performance
%measures such as prediction, evaluation amidst censored data is more
%complicated. First, we require
%fair performance measures that do not rely on the proportional hazards
%assumption, i.e. so-called
%model-free measures. Second, ``prediction" must be defined carefully
%as an outcome may not be observed
%for all $n$ subjects. Developing performance measures for survival
%data is an area of active research
%(cf. Begg et al., 2000; Steyerberg et al., 2010) and there is little
%concensus on which tools are preferred
%as each tool has its disadvantages.

%s4 ###
\section{Analysis of microarray data}\label{sec:results}

As described in Section~\ref{sec:intro}, the goal of our microarray
analysis is
to summarize the association of $p=1036$ gene expression levels with
survival time.
The combined data set from Harvard and Michigan consists of microarray
data for $n=200$ patients
with 46.5\% observations uncensored. We adopt boosting methods that
assume linearity in
the functional predictor $f(\bfX)=\sum_j \beta_j X_j$ and, hence,
coefficient estimates in the AFT model
have the interpretation of increase in average (logarithm) survival
time for a standard deviation (i.e., one unit)
increase in gene $X_j$ holding other factors fixed.

\begin{figure}

\includegraphics{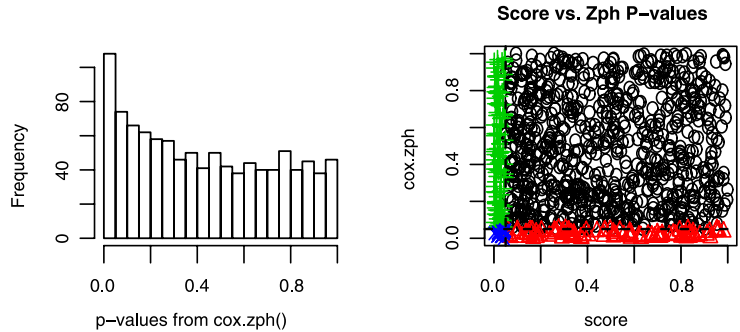}
%\vspace*{-5pt}
\caption{Results from diagnostic analyses on the microarray
data.}\label{fig:zph.pvals}
\end{figure}

First, we use heuristic methods to investigate the tenability of the
Cox PH model assumption in the microarray data.
All of our investigations are based on the \texttt{cox.zph} function
[Grambsch and Therneau (\citeyear{Grambsch1994})] in \texttt{R} which
tests the specific PH model assumption that the log relative hazard is
constant over time. Unless otherwise
stated, we
report the \texttt{cox.zph} global $p$-value for the full model. Similar
to Morris et al., we begin by fitting Cox models
one gene at a time. For each model, we record the $p$-value for the score
test as well as the $p$-value for the
diagnostic test; the results are displayed in Figure~\ref
{fig:zph.pvals}. The left panel in Figure~\ref{fig:zph.pvals}
is a histogram of the $p$-values for all 1036 diagnostic tests. A total
of 108 out of 1036 (10\%) diagnostic
tests rejected the null hypothesis at the nominal 0.05-level, that is,
about 5\% more than we would expect by chance
alone. If the PH model assumption were true marginally for each gene in
the microarray data, we would expect the
$p$-values in the left panel of Figure~\ref{fig:zph.pvals} to be
approximately uniformly distributed.
The right panel of Figure~\ref{fig:zph.pvals} displays the $p$-values
from the score test by diagnostic $p$-values from Cox regression
fits to the univariate models. We find that 16 of 108 genes (15\%) are
declared to be significantly related to
survival time at the nominal 0.05-level, but for which the diagnostic
test rejects the PH model assumption
of constant relative risk. Because the score test is not appropriate
for 108 of the univariate models, it
is unknown how\vadjust{\eject} many of the 16 genes are not significant at the nominal
level nor how many
of the 92 genes which are not declared significant by the score test
are, in fact, marginally associated with survival time.

%f1 ###

Because of multiple testing, it is known that drawing conclusions from
univariate models is naive. We further
investigated the tenability of the PH model assumption in multiple
regression models
using $p$-values from both the global test as well as individual
composite hypothesis tests
that the log relative hazard for a particular gene is constant given
other genes in
the model. The $p$-value for the global test did not drop below the
nominal 0.05-level until more than
80 genes entered the model. However, there was marginal evidence
($p\mbox{-value} = 0.06$) against the PH
assumption
using results from the composite tests with only the top three genes.
After one fits 20 or more of the
most significant genes into multiple regression Cox models, there is
always strong evidence ($p\mbox{-value} < 0.05$)
that at least one gene does not follow the PH model assumption given
other significant genes in the model.

We then investigated the sensitivity of the global diagnostic test to
the addition of noise variables. Here, we took
one of the 16 genes in Figure~\ref{fig:zph.pvals} that was associated
with survival time but did not satisfy the
PH model assumption in a univariate regression fit; the first one in
the data set was the 85th gene, so we chose
that one. We then fit a Cox PH model with the covariate vector
corresponding to this gene and an increasing
number of noise vectors, that is, $n$-dimensional vectors of standard
normal random variables.
The $p$-value for the global test was $0.03$, $0.09$, and $0.16$ with no,
one, and two noise vector(s),
respectively. The $p$-value for gene 85 from the Grambsch--Therneau
composite score test tended to stay closer to the
nominal level, although not always less than $0.05$. Hence, while our
application of
the diagnostic tool is imperfect for high
dimensional data analyses, there does seem to be some evidence contrary
to the PH model assumption
in joint analyses of the most significant genes identified through
univariate regression fits. A prudent approach is to
consider both the Cox PH and AFT models and we now describe results
from those analyses.

In Table~\ref{tab:camda.coef} we report the regression coefficients
from the final model after
5-fold CV for each of the PH and rank-based boostings. The third and
fifth columns present the ratio of a
given
coefficient over the largest coefficient in the active set. We see that
hazards regression selects a
nine-gene panel, while boosting the sum of pairwise differences selects
a 19-gene panel, with the
latter panel a proper superset of the former panel.
We see that the nine genes in the PH model are not the strongest nine
genes in the 19-gene panel.
Indeed, the sixth most significant gene in the PH model is the least
significant among all 19 in the
rank-based model. Moreover, the relative proportions indicate that the
panel composition is totally
different between the two methods.

%
%t1 ###
\begin{table}
\tabcolsep=0pt
\caption{Coefficients estimates for microarray data}\label{tab:camda.coef}
\begin{tabular*}{\textwidth}{@{\extracolsep{\fill}}ld{2.3}d{3.1}d{2.3}d{3.1}@{}}
\hline
\multicolumn{1}{l}{\textbf{Gene}} & \multicolumn{1}{c}{\textbf{PH}} & \multicolumn{1}{c}{\textbf{Rel. prop.}}
& \multicolumn{1}{c}{\textbf{Gehan}} & \multicolumn{1}{c@{}}{\textbf{Rel. prop.}} \\
\hline
Hs.34789 & -0.216 & 100.0 & 0.166 & 100.0 \\
Hs.146580 & 0.113 & 52.3 & -0.031 & 18.7 \\
Hs.119000 & 0.101 & 46.8 & -0.079 & 47.6 \\
Hs.406013 & 0.053 & 24.5 & -0.022 & 13.3 \\
Hs.407995 & 0.052 & 24.1 & -0.098 & 59.0 \\
Hs.75106 & -0.041 & 19.0 & 0.007 & 4.2 \\
Hs.174185 & -0.025 & 11.6 & 0.035 & 21.1 \\
Hs.2962 & 0.024 & 11.1 & -0.057 & 34.3 \\
Hs.2934 & 0.024 & 11.1 & -0.107 & 64.5 \\
Hs.28491 & \multicolumn{1}{c}{--} & 0 & -0.039 & 23.5 \\
Hs.82045 & \multicolumn{1}{c}{--} & 0 & 0.007 & 4.2 \\
Hs.576 & \multicolumn{1}{c}{--} & 0 & 0.021 & 12.7 \\
Hs.14231 & \multicolumn{1}{c}{--} & 0 & 0.070 & 42.2 \\
Hs.57301 & \multicolumn{1}{c}{--} & 0 & -0.038 & 22.9 \\
Hs.13046 & \multicolumn{1}{c}{--} & 0 & -0.023 & 13.9 \\
Hs.36602 & \multicolumn{1}{c}{--} & 0 & 0.014 & 8.4 \\
Hs.301132 & \multicolumn{1}{c}{--} & 0 & 0.058 & 34.9 \\
Hs.180107 & \multicolumn{1}{c}{--} & 0 & 0.035 & 21.1 \\
Hs.405945 & \multicolumn{1}{c}{--} & 0 & -0.036 & 21.7 \\
\hline
\end{tabular*}
\end{table}

We intended to display all three methods---PH, Gehan, and IPW---side-by-side in
Table~\ref{tab:camda.coef}, but the IPW active set included 94 genes
and the table could not
easily fit on one page.
In Table~\ref{tab:set.diff} we display the set differences among final
models by all three
methods. IPW shares only three genes in common with the PH model and
five genes in common
with the rank-based model. More than 95\% of the
genes in the IPW model \textit{do not} belong to either PH or
rank-based models. Thus, for
our microarray data analysis, the method proposed by Hothorn et al.
(\citeyear{Hothorn2006}) leads to very
different conclusions than the method proposed by Ridgeway (\citeyear
{Ridgeway1999}) and the one
proposed here.\looseness=1

%
%t2 ###
\begin{table}
\tabcolsep=0pt
\caption{Set differences (A--B) among PH, Gehan, and IPW
methods}\label{tab:set.diff}
\begin{tabular*}{250pt}{@{\extracolsep{\fill}}lcccc@{}}
\hline
  & & \multicolumn{3}{c@{}}{\textbf{Set B}}\\[-5pt]
  & & \multicolumn{3}{c@{}}{\hrulefill}\\
 \textbf{Set A} & \textbf{Total} & \textbf{PH} & \textbf{Gehan} & \textbf{IPW} \\
\hline
  PH & \hphantom{9}9 & \multicolumn{1}{c}{--} & \hphantom{8}0 & \hphantom{1}6 \\
  Gehan & 19 & 10 & \multicolumn{1}{c}{--} & 14 \\
  IPW & 94 & 91 & 89 & \multicolumn{1}{c@{}}{--}\\
\hline
\end{tabular*}
\end{table}

%s5 ###
\section{Comparative data analyses in large samples}\label{sec:compare}

In this section we provide two more real data examples where rank-based
ensembles lead
to different conclusions than PH or IPW ensembles. Unlike the
microarray data set, the following
data sets have many more observations than predictors. A consequence of
the large sample
size is that we may apply \texttt{cox.zph} directly to the entire data
set, as seen in the first
comparative data analysis. In the second analysis, we demonstrate that
differences between
rank-based and IPW ensembles transcend base learner and affects the
predictive scores
significantly.

%s5.1 ###
\subsection{Analysis of nursing home data}

From 1980--1982, the National Center for Health Services Research
conducted a
study to determine the effect of financial incentives on variation of
patient care
in nursing homes. In particular, 18 out of 36 nursing homes from San
Diego, California,
received higher per diem payments for accepting and admitting Medicaid patients
and additional bonuses when the patient's prognosis improved. The study
collected
data from an additional 18 control nursing homes where no financial
incentives were used.
A complete description is given in Morris, Norton and Zhou (\citeyear
{Morris1994}). The total sample size from all 36 nursing
homes is $n=1601$.

%
%t3 ###
\begin{table}
\tabcolsep=0pt
\tablewidth=210pt
\caption{Boosted coefficients estimates from the nursing home
data}\label{tab:nurse.home}
\centering
\begin{tabular*}{250pt}{@{\extracolsep{\fill}}ld{2.3}d{2.3}d{2.3}@{}}
\hline
& \multicolumn{1}{c}{\textbf{Cox}} & \multicolumn{1}{c}{\textbf{Gehan}} & \multicolumn{1}{c@{}}{\textbf{IPW}} \\
\hline
trt & -0.018 & 0.060 & 0.309 \\
age & -0.086 & 0.152 & 0.109 \\
sex & 0.165 & -0.283 & -0.129 \\
married & 0.061 & -0.066 & \multicolumn{1}{c@{}}{--} \\
h1 & \multicolumn{1}{c}{--} & \multicolumn{1}{c}{--} & \multicolumn{1}{c@{}}{--} \\
h2 & 0.098 & -0.215 & -0.160 \\
h3 & 0.157 & -0.291 & -0.212 \\
trt$*$age & 0.017 & \multicolumn{1}{c}{--} & \multicolumn{1}{c@{}}{--} \\
trt$*$sex & -0.015 & \multicolumn{1}{c}{--} & \multicolumn{1}{c@{}}{--} \\
trt$*$married & \multicolumn{1}{c}{--} & -0.027 & -0.124 \\
age$*$sex & 0.068 & -0.143 & -0.112 \\
age$*$married & 0.021 & - 0.005 & -0.048 \\
sex$*$married & \multicolumn{1}{c}{--} & \multicolumn{1}{c}{--} & \multicolumn{1}{c@{}}{--} \\
\hline
\end{tabular*}
\end{table}

Our data set consists of seven main effects and six 2-way interactions. The
main effects are treatment (trt), age, sex, marital status, and three
health status
indicators, ranging from the best health to the worst health. The 2-way
interactions are possible interactions among treatment, age, sex, and
marital status. This data set was previously analyzed by Fan and Li
(\citeyear{Fan2002}) using
the PH model. Using the Grambsch and Therneau (\citeyear
{Grambsch1994}) diagnostic test (\texttt{cox.zph} in \texttt{R})
for proportional hazards suggests that the PH model is inadequate
($\mbox{global }p\mbox{-value}=0.003$).
This does not prove the AFT model is correct but encourages us to look
beyond the PH model
in performing variable selection.

Table~\ref{tab:nurse.home} presents the results from our analysis of
the nursing home data. We
performed boosting in the PH and AFT models even though
our preliminary analysis indicated the inadequacy the former model.
Within the AFT model, we
performed coefficient ensembles using both the Gehan and IPW estimators
and noted that the two
models agree with the exception that IPW does not include the effect of
marital status in the final model.
A major difference between IPW and Gehan estimates is the magnitude of
the treatment effect:
three times the effect of age by the former method and less than half
the age effect in the latter method.
In comparing the Cox
and Gehan coefficient ensembles, we note that the inclusion and
exclusion of all seven main effects agree.
However, the presence or absence of all three interactions involving
treatment is reversed in the final
models. Both IPW and Gehan estimates exclude treatment-by-age and
treatment-by-sex interactions but
include the treatment-by-marital status interaction. The relative
magnitude of all two-way interactions is
modest to moderate with the age-by-sex interaction being strongest.

%s5.2 ###
\subsection{Blackbox methods on the Wisconsin PBC data}\label{sec:black}

Regression trees are the most common base procedure in the machine learning
community [Freund and Schapire (\citeyear
{Freund1996,Freund1997}); B\"{u}hlmann and Hothorn
(\citeyear{B2007})]
and nonparametric procedures are gaining popularity in applications
with complex data.
In this section we compare rank-based boosting to existing procedures for
survival outcomes using the Wisconsin Prognostic Breast Cancer (WPBC)
data set. The WPBC data set was contributed
by Street, Mangasarian and Wolberg (\citeyear{Street1995}) for
developing diagnostic models of breast cancer recurrence
and is available from the UCI repository for machine learning data bases.
The survival outcome is time to breast cancer recurrence and 30
predictors describe
features of cell nuclei taken from a digitized image of fine needle
aspirate of breast mass.
This data set was analyzed previously by B\"{u}hlmann and Hothorn
(\citeyear{B2007}) using inverse-probability
weighting (IPW) methods.

%
%f2 ###
\begin{figure}

\includegraphics{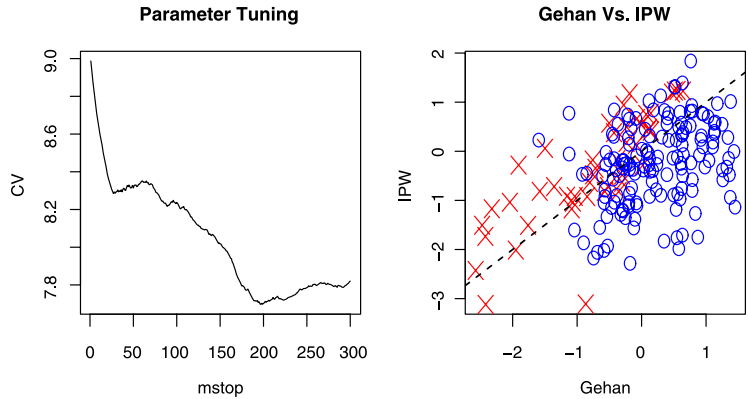}

\caption{Blackbox methods applied to the Wisconsin PBC data. Parameter
tuning for
Gehan loss is displayed in the left panel, while predictive scores
between Gehan and IPW losses
are illustrated in the right panel. Blue circles denote predicted
scores from uncensored
observations, while the red crosses denote censored
observations.}\label{fig:wpbc}
\end{figure}

Using regression trees as base learners, we boosted the following loss
functions:
IPW with $L_2$ loss, IPW with $L_1$
loss, and the sum of pairwise differences of absolute residuals (Gehan)
in \eqref{eq:LG}. Because
IPW boosting with~$L_1$ loss led to problems in identifiability, only
the results from $L_2$ loss are presented below.
Figure~\ref{fig:wpbc} summarizes output from the data analyses. The
left panel illustrates parameter tuning
for boosting the Gehan loss with regression trees as base learners and
shows that the optimal $m_{\mathrm{stop}}$ value
was 197 iterations when $\nu=0.1$. The right panel compares the
optimal fitted values $\widehat{f}(\bfX_i)$ from
blackbox Gehan and IPW methods using regression trees as base learners.
The blue circles denote predicted
scores from uncensored
observations, while the red crosses denote censored observations. We
note considerable disagreement
between the risk scores even though the same data, same AFT model, and
same boosting method are used in the statistical
learner---the only difference is in the loss function. Pearson's
correlation coefficient was only $r=0.45$. A total
of 66 (33\%) scores had different signs depending on the method; 52
(26\%) scores had absolute difference
greater than one unit and ten (5\%) scores were greater than two units
apart. Of the ten observations whose risk
scores were more than two units apart, nine observations were censored.
It is evident that boosting different
loss functions, particularly ones that embody different assumptions
about the underlying data, can result in
different estimated parameters and with potentially different conclusions.
We repeated\vadjust{\goodbreak} the analysis using smoothing splines as base
learners and found similar conclusions to the ones reported in
Figure~\ref{fig:wpbc}; thus, the
results from smoothing splines are not shown.

%s6 ###
\section{Simulations}\label{sec:sim}

%s6.1 ###
\subsection{Comparisons to partial likelihood}

We performed numerous simulation studies to compare coefficient
ensembles obtained from boosting
the partial likelihood to ensembles obtained from boosting the Gehan
loss function. We
simulated data according to the AFT model,
%
%e6.1 ###
\begin{equation}\label{eq:aft.sim}
\log T = \sum_{j=1}^d \beta_jX_j + \varepsilon,
\end{equation}
where $d=8$, the coefficient vector $\bfbeta=(3,3/2,0,0,2,0,0,0)\times
\kappa$, and the error
distribution was one of standard normal, extreme value (i.e., log
Weibull with unit shape), or a mixture distribution.
The mixture distribution was standard normal contaminated by a
Student's $t$ on three degrees of freedom
and contamination is controlled by a Bernoulli indicator with success
probability $0.2$. The predictors are
distributed multivariate normal with mean zero and covariance $\operatorname
{cov}(X_j,X_k)=(1/2)^{|j-k|}$. The
constant $\kappa$ controls the magnitude of the coefficient vector and
hence the signal-to-noise ratio;
here, we considered $\kappa$ equal
to $1/4$, $1/2$, $3/4$, and~$1$. A total of 100 Monte Carlo data sets
were computed
for each sample size of $n=60$, $80$, and $100$.

\begin{figure}[t!]

\includegraphics{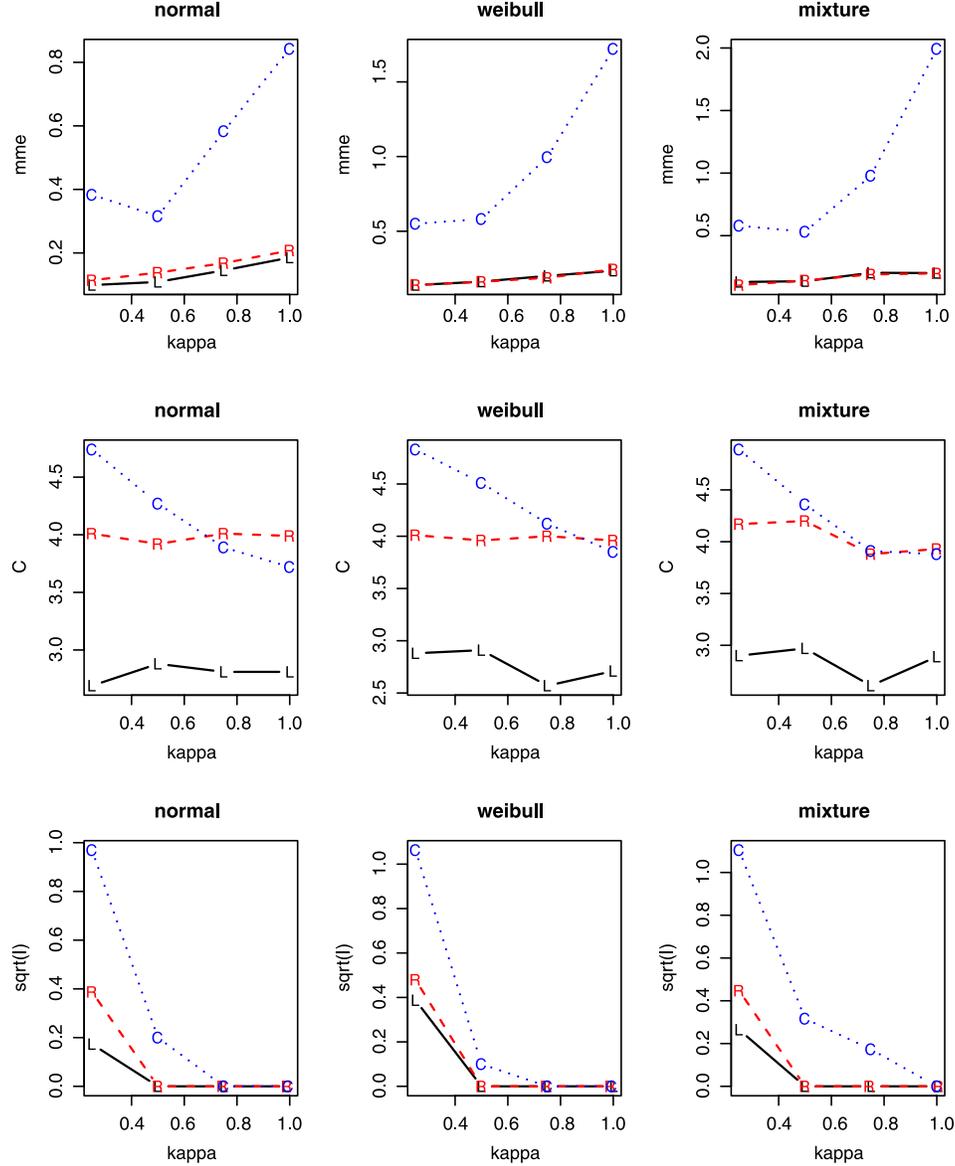}%
\vspace*{-5pt}
\caption{Simulation results in uncensored data. Coefficient ensembles displayed
for the Cox (C) PH model, rank-based (R) and $L_2$ (L)
boostings in a linear model (see text). Abscissa is expressed in
$\kappa$ (``kappa''), a multiplicative constant on
the coefficients in simulation studies, while ordinates refer to median
model error (mme), average
number of correct (C) and incorrect (I) zeros, with the last measure on
a square-root scale.}\label{fig:gehancox}\vspace*{-5pt}
\end{figure}

We evaluated estimators based on ubiquitous performance measures from
the variable selection literature:
model error (ME), mean squared error (MSE),
the average number of correct zeros (C), and the average number of
incorrect zeros (I). Regression
coefficient estimates from boosting the partial likelihood are
multiplied by minus one so that both estimators
are estimating the true coefficient vector $\bfbeta_0$ under extreme
value error distributions.
In the linear
model, prediction error is written as the sum of model error plus
noise, where model error is defined
$\mathrm{ME}=(\widehat{\bfbeta}-\bfbeta_0)^{\mathrm T}E(\bfX\bfX^{\mathrm
T})(\widehat{\bfbeta}-\bfbeta_0)$.
Although this performance measure is imperfect outside the linear
model, it
complements the other measures in a manner familiar to many
statisticians. The median ME (MME)
is presented in Figure~\ref{fig:gehancox}. The MSE is defined as ME
with the $d$-dimensional identity matrix
replacing the covariance matrix,
$E(\bfX\bfX^{\mathrm T})$. The definition of correct and incorrect zero
is straightforward; a~larger number is better
for the former measure, while a smaller number is better in the latter.

We performed coefficient selection and estimation in a variety of
simulation scenarios with and without censoring,
noting that censored data methods apply to complete data as well.
In Figure~\ref{fig:gehancox} we present the results for the uncensored
data case so that we might compare
our results to a routine implementation of $L_2$ boosting. So, the
three curves in Figure~\ref{fig:gehancox} refer
to $L_2$ (L) boosting, rank-based boosting (R), and boosting in the Cox
(C) PH model.

%
%f3 ###

Our simulation results indicate that as the magnitude of the true
coefficient vector increases, the ME and MSE for
$L_2$ and rank-based boosting remain about the same while that for the
PH model increases dramatically. This
result was true regardless of the error distribution, even when the PH
model assumptions were satisfied. However,
the gap in MSE between rank-based boosting and boosting the Cox model
decreases as the sample size $n$
increases. Interestingly, boosting in the Cox model has a tendency to
favor sparse models when effect sizes
are small and identifies models of similar complexity when effects
sizes are moderate to large. Again, there
was significant agreement in trends of model complexity across error
distributions. If one considers all four
performance measures together, then rank-based boosting is preferred.
If one dismisses ME and MSE as
unfair performance measures across PH and AFT models, the model
complexity via
rank-based boosting is less sensitive to the magnitude
of the effect size. In the presence of many small effects, the Cox
model will identify a larger number of
correct zeros but simultaneously overlook a larger proportion of true effects.

\begin{figure}

\includegraphics{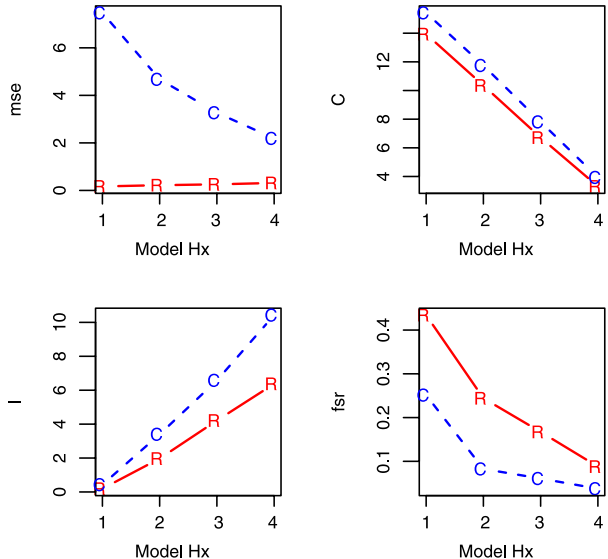}

\caption{Simulation results from comparing Cox (C) PH ensembles to
rank-based (R)
ensembles in uncensored data. Abscissa is labeled in terms of Models
H1--H4, with an increasing
number of nonzero coefficients for fixed theoretical $R^2$. Ordinates
refer to mean squared
error (mse), average number of correct (C) and incorrect (I) zeros, and
the false selection rate (fsr).}\label{fig:gcox.short}
\end{figure}

Next, we simulate censored data and no longer consider $L_2$ boosting.
We again simulate uncensored data
according to the linear model in \eqref{eq:aft.sim} with
autoregressive design and normal errors. True
regression coefficients are generated in two clusters according to the
following rule:
\begin{itemize}
\item for $h=1,\ldots ,4$, set initial coefficients $\beta
_{4+k,h}=\beta_{13+k,h}=(h-k)^2$, for $|k|<h$,
\item multiply initial coefficients by a constant to yield theoretical
$R^2=3/4$,
where theoretical $R^2$ for random design is
\[
R^2 = \frac{\bfbeta_0^{\mathrm T} E(\bfX\bfX^{\mathrm T}) \bfbeta_0}{\bfbeta
_0^{\mathrm T} E(\bfX\bfX^{\mathrm T}) \bfbeta_0 + \sigma^2},
\]
\end{itemize}
with $ \sigma$ the standard deviation of $\varepsilon_i$. Under this
simulation scenario, the signal strength
remains the same, while the proportion of active variables changes with
$h=1,\ldots ,4$.
Censored random variables were uniformly
distributed $\operatorname{Un}(0,\tau)$ to yield about 25\% censoring. As in
our earlier Monte Carlo studies, we monitor
the mean-squared error (MSE), average number of correct (C) and
incorrect (I) zeros. We also monitor
the average false selection rate (FSR), computed as the proportion of
unimportant variables relative to
the cardinality of the active set.
%To assess the relative performance of ensembles via Cox PH model versus
%rank-based ensembles, we use Harrell's (2001) C-index statistic, which
%measures the concordance
%of the linear predictor with the uncensored outcomes.
A~summary over 100 Monte Carlo data sets for
each of Model H1--H4 is displayed in Figure~\ref{fig:gcox.short}.

As in our earlier simulations, estimating regression coefficients under
an incorrect proportional hazards
assumption can lead to substantial bias. Hence, the MSE is much higher
using Cox PH compared to
rank-based estimation, but the bias decreases as the proportion of
active variables increases.
Remarkably, ensembles via Cox PH were mildly better than rank-based
ensembles in identifying
correct zeros, but the better performance came at a~price of
incorrectly setting important variables
to zero. Hence, PH ensembles tend to select models that are too sparse
under a
normal-theory linear model and many moderate effects are ignored completely.

%
%f4 ###

%s6.2 ###
\subsection{Comparisons to inverse-probability weighting}

We performed separate simulation studies to compare rank-based boosting and
inverse-probabi\-lity weighted (IPW) boosting [Hothorn et al. (\citeyear
{Hothorn2006})] in the AFT model.
We simulated data according to a similar AFT model used in earlier
simulation studies,
\[
\log T = \sum_{j=1}^d \beta_jX_j + \sigma\varepsilon,
\]
where the distribution of $\varepsilon$ follows one of standard
normal, log Weibull with unit shape,
or Student's $t$ on three degrees of freedom.
Two modeling differences are that we fixed the constant $\kappa=1$
and, hence, the coefficient vector is
$\bfbeta=(3,3/2,0,0,2,0,0,0)$ and varied the signal-to-noise ratio by
increasing the scale parameter
$\sigma$ from one-half to two in increments of $0.5$. We again used an
autoregressive design for the matrix of
predictors. The performance measures are identical to those discussed above.

\begin{figure}

\includegraphics{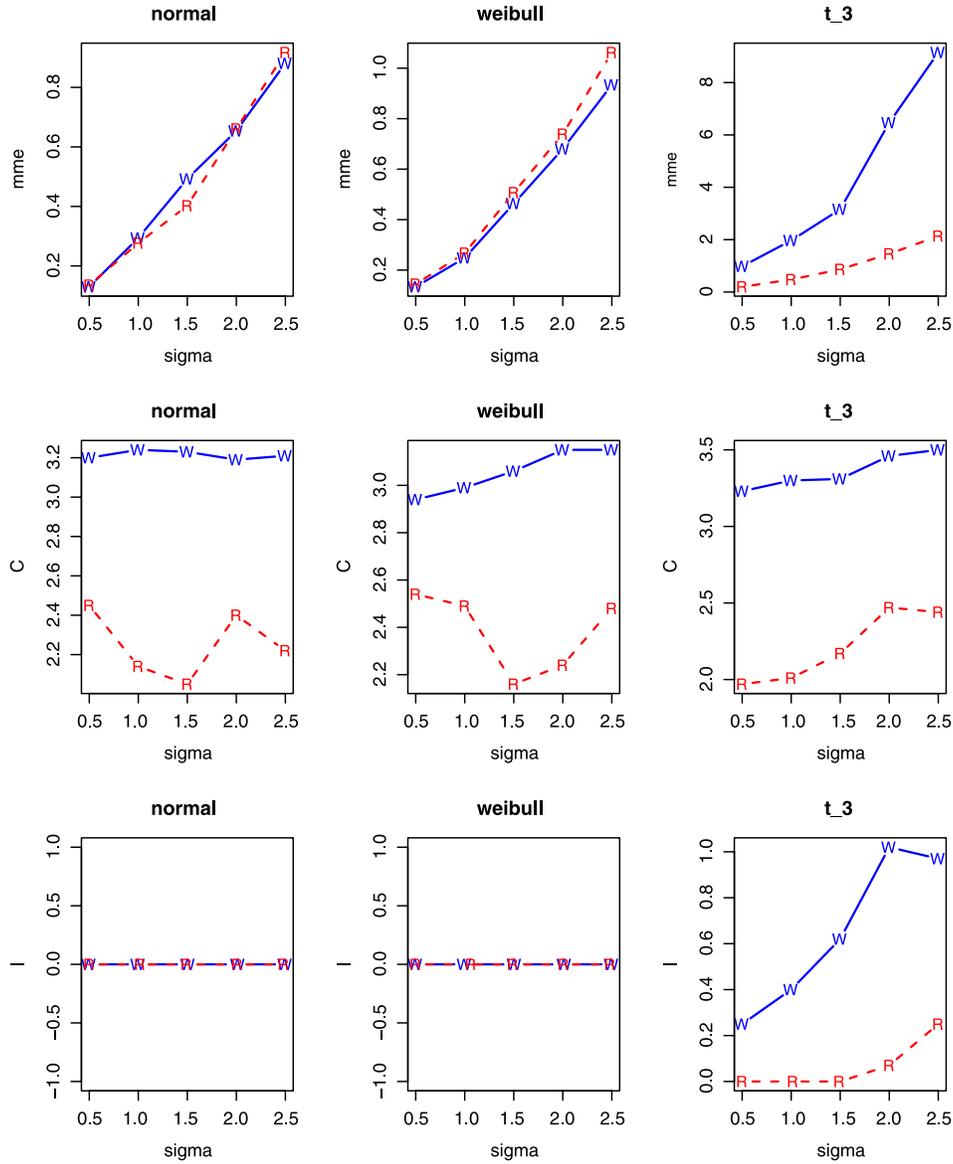}%
\vspace*{-5pt}
\caption{Simulation results from comparing inverse probability
weighting (W) versus rank-based (R) boosting via Gehan loss
when model assumptions are satisfied for both procedures. Abscissa
label is ``sigma,'' the scale
parameter on the errors in simulation studies; ordinate labels are
median model error (mme), average
number of correct (C) and incorrect (I) zeros.}\label{fig:gehanipw.1}\vspace*{-5pt}
\end{figure}

\begin{figure}

\includegraphics{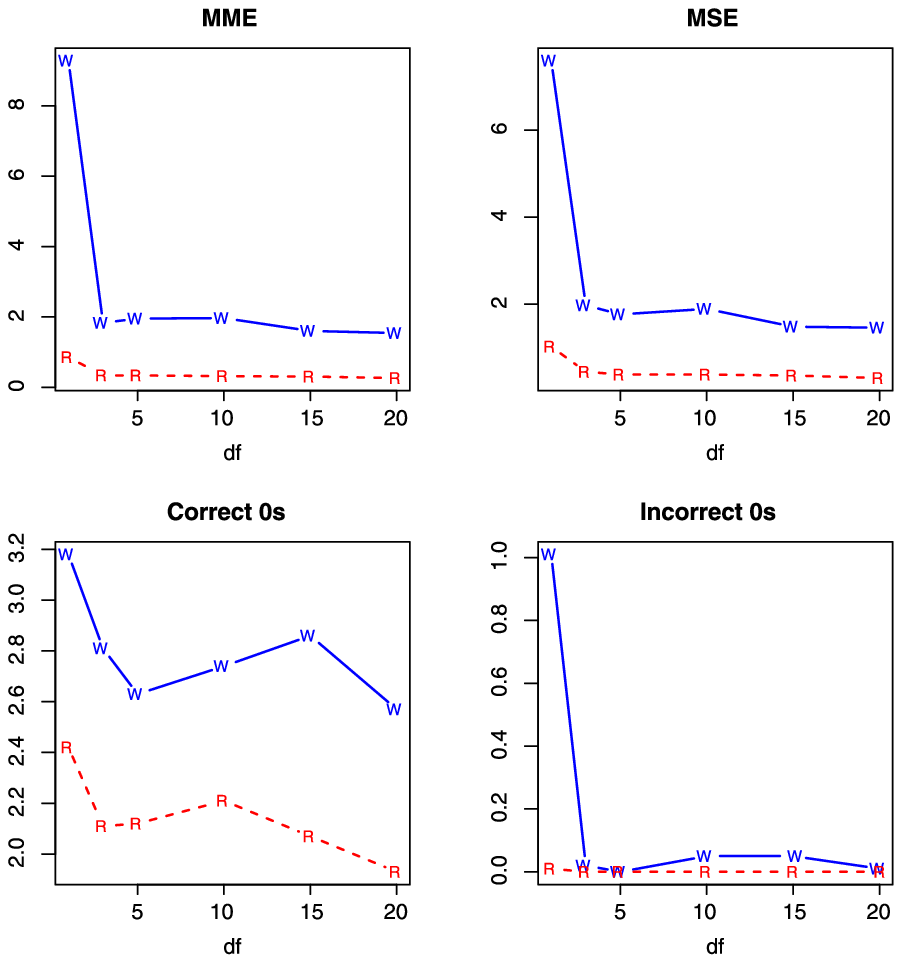}
\vspace*{-5pt}
\caption{Simulation results from comparing inverse probability
weighting (W) versus rank-based (R) boosting via Gehan loss
when IPW model assumptions are violated. Abscissa label is degrees of
freedom (df) of a Student's $t$
distribution from which the outcomes were generated; ordinate labels
are median model error (mme),
mean squared error, average
number of correct (C) and incorrect (I) zeros.}\label{fig:gehanipw.2}\vspace*{-5pt}
\end{figure}

The IPW boosting method by Hothorn et al. (\citeyear{Hothorn2006})
depends on correctly
modeling the (conditional)
censoring mechanism. But the procedure implemented in their \texttt
{mboost} \texttt{R} package makes the strong
assumption that censoring is independent of failure times. Here, we
simulate such data by generating
uniform censoring random variables, that is, $C\sim\operatorname{Un}(0,5)$.
Simulation results under the
independent censoring
assumption are provided in Figure~\ref{fig:gehanipw.1}. A weaker
assumption is to suppose
that censoring is conditionally independent of failure time given
covariates and we simulate such
data
through the model, $C=\bfbeta^{\mathrm T}\bfX+\operatorname{Un}(0,2)$.
Figure~\ref{fig:gehanipw.2} summarizes
simulation results under the latter modeling assumptions.
We report results for a~sample of size $n=60$
independent observations. In summary, the simulation scenarios are as follows:
\begin{itemize}
\item the error $\varepsilon$ is distributed as one of standard
normal, extreme value, or $t_3$; censoring is independent
of failure time and $C\sim\operatorname{Un}(0,5)$;
\item the error $\varepsilon$ is distributed as Student's $t$ on $r$
degrees of freedom, $r=1,3,5, 10,15,20$;
censoring is
conditionally independent of failure time given covariates and is modeled
$C=\bfbeta^{\mathrm T}\bfX+\operatorname{Un}(0,2)$.
\end{itemize}

When the stronger independent censoring assumptions are satisfied,
Figure~\ref{fig:gehanipw.1} suggests that IPW boosting
is a better procedure than rank-based boosting when the error
distribution has light tails but not when the
error distribution has heavy tails. Of course, the IPW methodology is
general and a more robust version of the IPW
may fix the deficiencies seen in the last column of the results in
Figure~\ref{fig:gehanipw.1} [Hothorn et al.
(\citeyear{Hothorn2006})]. However,
Figure~\ref{fig:gehanipw.2} suggests that excessive bias in IPW
coefficient ensembles is not
easily remedied by merely swapping loss functions. In Figure~\ref
{fig:gehanipw.2} we see that the ME and
MSE plateau around 1.9, while the rank-based Gehan procedure plateaus
at a value less than one-half.
Under the second simulation scenario, our results suggest that IPW
procedures produce estimates with excessive
bias and prefer models that are too simple. If a squared error loss
function is used to model lifetime data,
the IPW estimates will be sensitive to outlying lifetime values.

%
%f5 ###

%
%f6 ###

%s7 ###
\section{Remarks}\label{sec:remarks}

High-dimensional survival analysis is an important application
of the boosting machinery. In addition to placing weak restrictions
on the predictive function $f(\cdot)$, boosting applies to any well-defined
convex loss function. For survival outcomes, one's first inclination
may be to adopt
Cox's (\citeyear{Cox1972}) proportional hazards (PH) model and
estimate $f(\cdot)$ by minimizing
the negative log partial likelihood. It is known that the PH model
assumptions can
fail when $n\GG p$ and the problem persists for large-scale regression
problems too.
Diagnosing the veracity of the PH model assumptions in high dimensions
is an intriguing research
problem and beyond the scope of the current manuscript.
When the proportional hazards assumption is inadequate or false,
the accelerated failure time (AFT) model may be a reasonable
alternative. In Section~\ref{sec:sim}
we illustrated the effects on variable selection when the PH model is
misspecified but
the AFT model is correctly specified.
Head-to-head comparisons of coefficient ensembles via hazards versus
linear modeling suggest that the former identifies a slightly higher
proportion of correct zeros and has
a lower proportion of unimportant variables in final models even when
the PH model assumption
is incorrect. At the same time, in addition to expected excessive bias
in misspecified models, we find that PH
ensembles tend to select models that are too sparse and, hence, ignore
many moderate effects.

Hothorn et al. (\citeyear{Hothorn2006}) discussed boosting survival
data in the semi-parame\-tric
AFT model via inverse probability weighting (IPW).
The coefficient ensembles by Hothorn et al. (\citeyear{Hothorn2006})
are built on a theory of
inverse probability weighting (IPW), a powerful technique for general
missing data problems [van der Laan and Robins (\citeyear{Laan2003});
Tsiatis (\citeyear{Tsiatis2006})].
The idea is to model the censoring
mechanism as a function of covariates and then weight the uncensored
data by the reciprocal of the estimated
``complete case'' probability. If the censoring mechanism does not
depend on covariates,
then the modeling is accomplished nonparametrically via Kaplan--Meier
estimation.
In contrast, the ensemble methods proposed here are based on
minimizing rank-based dispersion criteria and do not require modeling the
censoring mechanism. Both IPW and rank-based ensemble
methods minimize convex loss functions and fit neatly into the boosting
template; however, rank-based
methods operate under less stringent conditions. Although second-stage
modeling can improve the simple
inverse-weighting method proposed by Hothorn et al. (\citeyear
{Hothorn2006}), the secondary models would be difficult to verify and
computational details left to the user. Our simulation studies
indicated that the boosting method
by Hothorn et al. (\citeyear{Hothorn2006}) is as good or better than
rank-based boosting in
special cases but not in general.

In conclusion, our analyses and simulation studies indicate that each
of PH, rank-based, and IPW
survival ensembles
can exhibit the best and worst learning behavior depending on the
model, the data, and
how one evaluates model performance. We were surprised that boosting in
Cox's PH model
performed as well as it did even in a misspecified linear model with
normal errors. But even in this
simple simulation scenario, we were able to replicate the types of
difference betweeen PH and rank-based
ensembles and potential underfitting behavior similar to what we saw in
our microarray data set.
We feel that rank-based survival ensembles have merit and will provide
scientists with
a strong tool for investigating large data sets in a wide variety of
settings.

%We used Harrell's (2001) concordance measure to assess model
%performance when the PH model
%assumption is and is not satisfied. In many of the scenarios that we
%investigated, we found
%the average concordance of PH ensembles was comparable to rank-based
%ensembles.
%Harrell's measure is simple in that censored observations
%contribute nothing to the measure because the outcome is missing. This
%simplicity is also a
%criticism and one reason why authors attempt to improve it (Begg et
%al., 2000; Steyerberg et al., 2010).
%It is unclear
%why we failed to show that rank-based ensembles led to improved
%concordance over
%PH ensembles even when the PH assumption was incorrect. Two possible
%reasons include
%the insensitivity of Harrell's statistic or the boosting procedure
%itself. Investigating the details of
%this mystery and developing improved model-free performance measures
%are subjects of
%future research. It is also important to remember that a statistical
%learner that exceeds by one
%criterion is likely to be deficient in another.

%suskaldyti doi

\section*{Acknowledgments}
We thank Editor Kafadar, an Associate Editor, and two referees for
constructive comments which improved the paper.

\printaddresses


\begin{thebibliography}{99}

%Maximum likelihood identification of Gaussian autoregressive moving
%average models.

%J.~P.} (2004)
% Metagenes and molecular pattern discovery using matrix factorization.
% \textit{PNAS} {\bf101}, 4164--4169.

%b1 ###
\bibitem[\protect\citeauthoryear{}{2004}]{Blanchard2004}
\textsc{Blanchard, G., Lugosi, G.} and \textsc{Vayatis, N.} (2004).
On the rate of regularized boosting classifiers.
\textit{J. Mach. Learn. Res.} \textbf{4} 861--894.
\MR{2076000}

\bibitem[\protect\citeauthoryear{Breiman}{1998}]{Br1998}
\textsc{Breiman}, L. (1998).
 Arcing classifiers (with discussion).  \textit{Ann. Statist.}  \textbf{26}
801--849.


%b3 ###
\bibitem[\protect\citeauthoryear{}{2007}]{B2007}
\textsc{B\"{u}hlmann, P.} and \textsc{Hothorn, T.} (2007).
Boosting algorithms: Regularization, prediction and model fitting (with
discussion).
\textit{Statist. Sci.} \textbf{22} 477--522.
\MR{2420454}

%b2 ###
\bibitem[\protect\citeauthoryear{}{2003}]{B2003}
\textsc{B\"{u}hlmann, P.} and \textsc{Yu, B.} (2003).
Boosting with the $L_2$ loss: Regression and classification.
\textit{J.~Amer. Statist. Assoc.} \textbf{98} 324--339.
\MR{1995709}

% Rank estimation in partial linear model with censored data.
% \textit{Statistica Sinica} \textbf{15} 767--779.

%b4 ###
\bibitem[\protect\citeauthoryear{}{1972}]{Cox1972}
\textsc{Cox, D.~R.} (1972).
Regression models and life-tables (with discussion).
\textit{J. Roy. Statist. Soc. Ser. B} \textbf{34} 187--202.
\MR{0341758}

%b5 ###
\bibitem[\protect\citeauthoryear{}{1984}]{Cox1984}
\textsc{Cox, D.~R.} and \textsc{Oakes, D.} (1984).
\textit{Analysis of Survival Data}. Chapman \& Hall, London.
\MR{0751780}

%Least angle regression (with Discussion).

%b6 ###
\bibitem[\protect\citeauthoryear{}{2002}]{Fan2002}
\textsc{Fan, J.} and \textsc{Li, R.} (2002).
Variable selection for Cox's proportional hazards model and frailty model.
\textit{Ann. Statist.} \textbf{30} 74--99.
\MR{1892656}

%New York: Wiley.

%b7 ###
\bibitem[\protect\citeauthoryear{}{1996}]{Freund1996}
\textsc{Freund, Y.} and \textsc{Schapire, R.} (1996).
Experiments with a new boosting algorithm. In   \textit{Proceedings of the
13th International Conference on Machine Learning Theory}.
Morgan Kaufmann, San Francisco, CA.

%b8 ###
\bibitem[\protect\citeauthoryear{}{1997}]{Freund1997}
\textsc{Freund, Y.} and \textsc{Schapire, R.} (1997).
A decision-theoretic generalization on on-line learning and an
application to boosting.
\textit{J. Comput. System Sci.} \textbf{55} 119--139.
\MR{1473055}

%b9 ###
\bibitem[\protect\citeauthoryear{}{2001}]{Friedman2001}
\textsc{Friedman, J.} (2001).
Greedy function approximation: A gradient boosting machine.
\textit{Ann. Statist.} \textbf{29} 1189--1232.
\MR{1873328}

%b10 ###
\bibitem[\protect\citeauthoryear{}{2000}]{Friedman2000}
\textsc{Friedman, J., Hastie, T.} and \textsc{Tibshirani, R.} (2000).
Additive logistic regression: A~statistical view of boosting (with
discussion).
\textit{Ann. Statist.} \textbf{28} 337--407.
\MR{1790002}

%Gradient directed regularization.
%Technical report, Stanford University, Department of Statistics.

%Pathwise coordinate optimization.

%b11 ###
\bibitem[\protect\citeauthoryear{}{1965}]{Gehan1965}
\textsc{Gehan, E.~A.} (1965).
A generalized Wilcoxon test for comparing arbitrarily single-censored samples.
\textit{Biometrika} \textbf{52} 203--223.
\MR{0207130}

%b12 ###
\bibitem[\protect\citeauthoryear{}{1994}]{Grambsch1994}
\textsc{Grambsch, P.~M.} and \textsc{Therneau, T.~M.} (1994).
Proportional hazards tests and diagnostics based on weighted residuals.
\textit{Biometrika} \textbf{81} 515--526.
\MR{1311094}

%%b13 ###
%Stratified gradient boosting for fast training of conditional random fields.
% In \textit{5th International Workshop of Mining and Learning
%with Graphs}.
%Universita degli Studi di Firenze, Florence, Tuscany, Italy.
%
%%b14 ###
%Estimates of location based on rank tests.

%b15 ###
\bibitem[\protect\citeauthoryear{}{2007}]{Hothorn2007}
\textsc{Hothorn, T.} and \textsc{B\"{u}hlmann, P.} (2007).
Mboost: Model-based boosting.
R~package version 0.5-8. Available at \url{http://CRAN.R-project.org/}.

%b16 ###
\bibitem[\protect\citeauthoryear{}{2006}]{Hothorn2006}
\textsc{Hothorn, T.}, \textsc{B\"{u}hlmann, P.}, \textsc{Dudoit,
S.}, \textsc{Molinaro, A.} and
\textsc{van der Laan, M.} (2006).
Survival ensembles.
\textit{Biostatistics} \textbf{7} 355--373.

%b17 ###
\bibitem[\protect\citeauthoryear{}{1972}]{Jaeckel1972}
\textsc{Jaeckel, L.~A.} (1972).
Estimating regression coefficients by minimizing the dispersion of the
residuals.
\textit{Ann. Math. Statist.} \textbf{43} 1449--1458.
\MR{0348930}

%b18 ###
\bibitem[\protect\citeauthoryear{}{2008}]{Johnson2008}
\textsc{Johnson, B.~A., Lin, D.~Y.} and \textsc{Zeng, D.} (2008).
Penalized estimating functions and variable selection in semiparametric
regression models.
\textit{J. Amer. Statist. Assoc.} \textbf{103} 672--680.
\MR{2435469}

%b19 ###
\bibitem[\protect\citeauthoryear{}{2002}]{Kalbfleisch2002}
\textsc{Kalbfleisch, J.~D.} and \textsc{Prentice, R.~L.} (2002).
\textit{The Statistical Analysis of Failure Time Data.}
Wiley, New York.
\MR{1924807}

%b20 ###
\bibitem[\protect\citeauthoryear{}{1993}]{Lin1993}
\textsc{Lin, D.~Y., Wei, L.~J.} and \textsc{Ying, Z.} (1993).
Checking the Cox model with cumulative sums of martingale-based residuals.
\textit{Biometrika} \textbf{80} 557--581.
\MR{1248021}

%%b21 ###
%Boosting method for nonlinear transformation models with censored
%survival data.

%b22 ###
\bibitem[\protect\citeauthoryear{}{1994}]{Morris1994}
\textsc{Morris, C.~N.}, \textsc{Norton, E.~C.} and \textsc{Zhou, X.~H.}
(1994).
Parametric duration analysis of nursing home usage.
In \textit{Case Studies in Biometry}
(N. Lange, L. Ryan, L. Biilard, D.~Brillinger, L. Conquest and J.
Greenhouse, eds.)
231--248.
Wiley, New York.

%b23 ###
\bibitem[\protect\citeauthoryear{}{2005}]{Morris2005}
\textsc{Morris, J.~S., Yin, G., Baggerly, K.~A., Wu, C.} and \textsc
{Zhang, L.}
(2005).
Pooling information across different studies and oligonucleotide microarray
chip types to identify prognostic genes for lung cancer. In
\textit{Methods of Microarrray
Data Analysis IV} (J.~S.~Shoemaker and
S.~M. Lin, eds.). Springer, New York.

% Bi-cross-validation of the SVD and the nonnegative matrix
%factorization.
% \textit{Ann. Appl. Statist.} \textbf{3} 564--594.

%b24 ###
\bibitem[\protect\citeauthoryear{}{1978}]{Prentice1978}
\textsc{Prentice, R.~L.} (1978).
Linear rank tests with right-censored data.
\textit{Biometrika} \textbf{65} 167--179.
\MR{0497517}

%b25 ###
\bibitem[\protect\citeauthoryear{}{1994}]{Reid1994}
\textsc{Reid, N.} (1994).
A conversation with Sir David Cox.
\textit{Statist. Sci.} \textbf{9} 439--455.
\MR{1325436}

%b26 ###
\bibitem[\protect\citeauthoryear{}{1999}]{Ridgeway1999}
\textsc{Ridgeway, G.} (1999).
The state of boosting.
\textit{Computing Science and Statistics} \textbf{31} 172--181.

%%b27 ###
%Flexible boosting of accelerated failure time models.

%b28 ###
\bibitem[\protect\citeauthoryear{}{1982}]{Schoenfeld1982}
\textsc{Schoenfeld, D.} (1982).
Partial residuals for the proportional hazards regression model.
\textit{Biometrika} \textbf{69} 239--241.

%b29 ###
\bibitem[\protect\citeauthoryear{}{1995}]{Street1995}
\textsc{Street, W.~N., Mangasarian, O.~L.} and \textsc{Wolberg, W.~H.}
(1995).
An inductive learning approach to prognostic prediction. In
\textit{Proceedings
of the 12th International Conference on Machine Learning}.
Morgan Kaufmann, San Francisco, CA.

%b30 ###
\bibitem[\protect\citeauthoryear{}{2000}]{Therneau2000}
\textsc{Therneau, T.~M.} and \textsc{Grambsch, P.~M.} (2000).
\textit{Modeling Survival Data: Extending the Cox Model}.
Springer, New York.
\MR{1774977}

%b31 ###
\bibitem[\protect\citeauthoryear{}{1990}]{Tsiatis1990}
\textsc{Tsiatis, A.~A.} (1990).
Estimating regression parameters using linear rank tests for censored data.
\textit{Ann. Statist.} \textbf{18} 354--372.
\MR{1041397}

%b32 ###
\bibitem[\protect\citeauthoryear{}{2006}]{Tsiatis2006}
\textsc{Tsiatis, A.~A.} (2006).
\textit{Semiparametric Theory and Missing Data}.
Springer, New York.
\MR{2233926}

%b33 ###
\bibitem[\protect\citeauthoryear{}{2003}]{Laan2003}
\textsc{van der Laan, M.~J.} and \textsc{Robins, J.~M.} (2003).
\textit{Unified Methods for Censored Longitudinal Data and Causality}.
Springer, New York.
\MR{1958123}

% A large sample study of rank estimation for censored regression data.
% \textit{Ann. Statist.} \textbf{21} 76--99.

%On the degrees of freedom of lasso.

%b34 ###
\bibitem[\protect\citeauthoryear{}{2005}]{Zhang2005}
\textsc{Zhang, T.} and \textsc{Yu, B.} (2005).
Boosting with early stopping: Convergence and consistency.
\textit{Ann. Statist.} \textbf{33} 1538--1579.
\MR{2166555}
\end{thebibliography}
\end{document}